# THE HIDDEN COST OF ACCOMMODATING CROWDFUNDER PRIVACY PREFERENCES: A RANDOMIZED FIELD EXPERIMENT

GORDON BURTCH[*]     ANINDYA GHOSE[**]     SUNIL WATTAL[***][1]

August 12[th], 2014

*Forthcoming,* **Management Science**

[*] UMN Carlson School of Management. gburtch@umn.edu

[**] NYU Stern School of Business. aghose@stern.nyu.edu

[***] Temple Fox School of Business. swattal@temple.edu

[1] The authors are grateful to reviewers and participants at the 2013 Workshop on Information Systems and Economics (WISE), the 2013 INFORMS Annual Meeting, the 2014 Academy of Management (AoM) Annual Meeting, the 2014 NBER Summer Institute on the Economics of Digitization and IT and the 2014 ZEW Conference on the Economics of ICT. The authors also thank Sean Taylor for providing helpful comments. Anindya Ghose acknowledges funding from NSF CAREER Award IIS-0643847.

**Abstract**

Online crowdfunding has received a great deal of attention from entrepreneurs and policymakers as a promising avenue to fostering entrepreneurship and innovation. A notable aspect of this shift from an offline to an online setting is that it brings increased visibility and traceability of transactions. Many crowdfunding platforms therefore provide mechanisms that enable a campaign contributor to conceal his or her identity or contribution amount from peers. We study the impact of these information (privacy) control mechanisms on crowdfunder behavior. Employing a randomized experiment at one of the world's largest online crowdfunding platforms, we find evidence of both positive (e.g., comfort) and negative (e.g., privacy priming) causal effects. We find that reducing access to information controls induces a net increase in fundraising, yet this outcome results from two competing influences – treatment increases willingness to engage with the platform (a 4.9% increase in the probability of contribution) and simultaneously decreases the average contribution (a $5.81 decline). This decline derives from a publicity effect, wherein contributors respond to a lack of privacy by tempering extreme contributions. We unravel the causal mechanisms that drive the results and discuss the implications of our findings for the design of online platforms.

# THE HIDDEN COST OF ACCOMMODATING CROWDFUNDER PRIVACY PREFERENCES: A RANDOMIZED FIELD EXPERIMENT

## 1. Introduction

Over the last eight years, a growing proportion of the venture finance gap has been filled by novel funding mechanisms. Online crowdfunding, defined as "a collective effort by individuals who network and pool their money together, usually via the Internet, to invest in or support the efforts of others" (Ordanini et al. 2010), has received a great deal of attention from entrepreneurs and policymakers as a promising avenue to fostering entrepreneurship and innovation.

One of the primary hurdles typically faced by new entrepreneurs is the identification and sourcing of capital (Wetzel 1987). Crowdfunding simplifies this process by providing entrepreneurs with broader reach and visibility (Agarwal et al. 2013; Kim and Hann 2013). However, a notable implication of shifting the fundraising process online is the increased visibility and traceability of transactions. Most crowdfunding platforms maintain a public record of all transactions, including information about a contributor's identity, the amount of their contribution and the campaign they chose to support. Perhaps cognizant of the potential that some crowdfunders may shy away from scrutiny, while others may seek it out, many crowdfunding platforms now provide users with transaction-level information controls that enable concealment (revelation) of identity or contribution amounts, at the contributor's discretion.

Ostensibly, providing crowdfunders with the ability to determine the visibility of their contributions to peers should increase their level of satisfaction, and thus their willingness to transact, resulting in increased fundraising. A large number of studies in recent years support this logic. Scholars have noted the growing prevalence of privacy concerns amongst consumers (Goldfarb and Tucker 2012), and have demonstrated the positive effects of privacy assurances, policies and seals on user information sharing and product purchase (Hui et al. 2007; Tsai et al. 2011). At the same time, a number of studies have demonstrated the value of social recognition and reputational gains as drivers of user contributions to online communities (Wasko and Faraj 2005; Zhang and Zhu 2011).

However, providing users with information control mechanisms can also be costly. Recent work suggests users may ignore these features if they perceive that they are inflexible, difficult to understand or a challenge to use (Das and Kramer 2013; Sleeper et al. 2013), potentially opting not to transact at all. It has also been shown that prompting individuals with questions about scrutiny or their privacy can elicit privacy concerns, via priming effects (John et al. 2011; Joinson et al. 2008; Tucker 2014). This, in turn, can have a negative influence on users' willingness to engage with a purveyor, platform or other users.

We therefore set out to understand the impact that transaction-level information controls have on crowdfunders' willingness to contribute, as well as their subsequent behavior, conditional on conversion. More specifically, ***we seek to identify and quantify the causal relationship between a crowdfunding platform's information control features and contributors' willingness to transact. Further, we look to understand any associated shifts in crowdfunder behavior, conditional on transaction.***

Evaluating the impact of information control provision on user behavior is inherently challenging because of various biases associated with observational and survey-based attempts to evaluate privacy sensitive individuals who frequently are, by definition, unwilling to be scrutinized or profiled. Moreover, concerns about privacy may not be accurately reflected in interview or survey-based settings because of the gap between consumers' claimed privacy concerns and their actual behavior in response to those concerns (Strandburg 2005). Experimental subjects expect a researcher to collect their information and they are unlikely to have concerns about receiving unwanted solicitations from third party organizations or individuals down the line, because standard data collection policies in experimental protocols prohibit the sharing of data without consent (Wattal et al. 2012).

Meanwhile, observational studies are confronted with their own, comparable issues. Researchers must contend with a lack of available data, as subjects strive to conceal their actions. Further, issues of endogeneity stemming from self-sorting and self-selection (Heckman 1979) are also likely to arise from any changes in user privacy conditions. To clarify, consider the example of the privacy sensitive consumer. Such consumers may opt to exit a marketplace entirely following, for example, the removal of a privacy assurance. Unless this selection effect were to somehow be accounted for explicitly, either in the data or through estimation techniques, it would be impossible to draw valid, generalizable conclusions about the effect that such a change had on user behavior. While various econometric techniques are available to address these issues, each is heavily laden with assumptions. Further, data based adjustments are often challenging if not impossible to implement, as subjects who do not participate will often go unobserved.

Fortunately, web-based experimentation with impression- or session-level data can alleviate many of these concerns. We partnered directly with the purveyor of a leading global online crowdfunding platform to design and execute a randomized control trial, unbeknownst to the subjects under study (i.e., website visitors). Subjects in our sample were thus observed while making real-life decisions, with real economic consequences. Our results are therefore not subject to the reporting biases inherent in survey research of privacy issues, nor are they subject to issues of self-selection, because we observe subjects even when they choose not to transact.

We randomly manipulate the presentation timing of an information control question, displaying it either before or after the completion of payment. This intervention allows us to understand what impact information control features have on users' willingness to engage with the website, in terms of whether they contribute to crowdfunding campaigns (willingness to transact) and, given contribution, their contribution amounts.

We found, counter to intuition, that delaying the presentation of information controls drove a 4.9% increase in users' probability of completing a transaction. At the same time, conditional on transacting, the dollar amount of the average campaign contribution declined (by $5.81) with treatment. Fortunately for the purveyor of this platform, the increase in the rate of participation was sufficient to offset the decline in average contributions, resulting in an immediate net benefit from the intervention. Accordingly, the purveyor has since adopted the post-payment setup on a permanent basis.

Our subsequent analyses indicate that the treatment reduced the variance of contribution amounts, with an asymmetrically stronger effect on large contributions. That is, our treatment reduced the prevalence of both large and small contributions, though the decline in large contributions was more pronounced. We submit that this occurred because contributors in the post-payment condition, having reduced access to privacy controls, perceived greater publicity for their actions. Accordingly, they regressed toward the mean to avoid drawing unwanted attention (e.g., unsolicited requests from other crowdfunding campaigns). This result provides empirical evidence of the impact of publicity on individuals' behavior, which has seen theoretical consideration in the prior literature on monetary donations to public goods (Daughety and Reinganum 2010), and which has also been demonstrated in regard to other types of online contributions, such as user-generated content in the form of restaurant reviews (Wang 2010). This implies that a careful balance must be maintained between users' privacy concerns and the behavior that can ensue from accommodating those concerns.

This work contributes to the growing literature on crowdfunding (Agarwal et al. 2013; Burtch et al. 2013a). Studies have looked at various drivers of campaign fundraising outcomes, including pitch framing and information disclosure (Ahlers et al. 2012; Herzenstein et al. 2011b), fundraisers' social networks (Lin et al. 2013; Mollick 2014), geography and culture (Agarwal et al. 2011; Burtch et al. 2014; Lin and Viswanathan 2013). However, perhaps the most common subject of study has been peer influence amongst contributors (Burtch 2011; Burtch et al. 2013b; Herzenstein et al. 2011a; Kim and Viswanathan 2013; Zhang and Liu 2012). Bearing in mind that peer influence depends upon the visibility of peers' actions, our work considers the underlying context and mechanisms that enable those effects.

Our work also builds on the literatures dealing with privacy and reputation. In that we consider the dual, potentially countervailing impacts of privacy feature provision on user behavior: the effect on conversion rates and conditional contributions, which to our knowledge have not previously been examined in tandem. Lastly, our work contributes to the literature on anonymity in charitable contribution. A number of studies in recent years have noted the role of perception management and social image in pro-social behavior (Andreoni and Bernheim 2009; Daughety and Reinganum 2010). Our results indicate that these types of concerns similarly play into crowdfunder behavior, which in turn speaks to the presence of altruistic motives in online crowdfunding markets.

Before presenting our experimental design, analysis and results, we offer a review of the literature dealing with crowdfunding, in order to provide the reader with an understanding of the role that privacy or information controls may play on these platforms. We then provide a review of the literature dealing with online privacy and information hiding, before delving into our methods.

## 2. Methods: Randomized Experiment

### 2.1 Study Context

Our experiment was conducted at one of the largest global reward-based crowdfunding platforms, which enables anyone to raise money for a project or venture. The marketplace attracts upwards of 200,000 visitors per day, and facilitates millions of dollars in contributions each month. The platform has attracted over 1 million users since 2008, from more than 200 countries.

The platform allows fundraising for any purpose. When campaign owners first submit their project, they are required to specify how the money will be used, rewards that contributors can claim, the target amount to be raised, the number of days the fundraiser will run for, and the funding format (keep what you raise versus a provision point mechanisms / threshold fundraiser).

Campaigns are presented to website visitors in order of popularity. Popularity is measured algorithmically by the platform operator, based on a variety of factors, including organizer effort, fundraising progress, media coverage, etc., The homepage highlights new campaigns and those that are ending soon. A visitor can also filter ongoing campaigns by location, proximity ("near me") or category[2].

Once an individual decides to contribute, they must specify how many dollars they would like to supply. Next, contributors provide an e-mail address and, if a reward is being claimed, a shipping address. At this point, in our control condition, the contributor is presented with a question about how they want their contribution record to appear to website visitors. The contributor can conceal either their identity or the amount of the contribution (but not both)[3]. Importantly, a contributor's identity and amount will always be visible to the campaign organizer and platform operator; this information control prompt only masks details from a contributor's peers.

### 2.2 Experimental Design

Figure 1 presents a design mockup of the information control question that is posed to users during the course of contribution. The user is asked to specify which pieces of information about the contribution they would like to display publicly[4]. Our experimental treatment imposes a delay in the presentation of this question, from before payment to after payment. This treatment mimics removal of the mechanism from the platform, in a watered down form. This treatment allows us assess the economic impact of providing information controls, both in terms of users' willingness to transact and contribution amounts, conditional on transaction. Ultimately, we aim to assess whether these mechanisms deliver a net benefit or detriment to campaign fundraisers and the platform operator.

---

[2] The campaign organizer (rather than the marketplace purveyor) determines the campaign category. As such, there are no strict rules around the assignment of categories, thus these groupings are fuzzy and may overlap.

[3] Information-hiding mechanisms of this sort are relatively common in online crowdfunding. Some other prominent platforms that employ these features include GoFundMe.com, GiveForward.com, and CrowdRise.com.

> Your contribution will currently appear to everyone as "John Doe - $20."
> Would you like to change the appearance?
>
> ⦿ **Name & Amount:** "John Doe – $20"
> ⦿ **Name Only:** "John Doe – Undisclosed"
> ⦿ **Amount Only:** "Anonymous – $20"

**Figure 1. Privacy Control Prompt**

As noted above, in the pre-payment (control) condition, the information control question is presented to the user just prior to payment. In the post-payment (treatment) condition, the mechanism is not presented until after payment has been completed. Figure 2 provides a visual comparison of the experimental flow experienced by subjects in our treatment and control conditions.

The timing of the information control prompt (i.e., before vs. after payment) may have two foreseeable, countervailing impacts on user behavior. On the one hand, placing the mechanism after payment may reduce any potential privacy or scrutiny priming effects, as the user is not prompted to consider these issues before making their payment. In turn, this effect could be expected to increase conversion rates. On the other hand, delaying presentation might reduce willingness to transact if users already have privacy or scrutiny in mind – e.g., privacy sensitive individuals. This may reduce willingness to engage or transact. Because of these competing effects, it is not immediately clear what impact our treatment will have on fundraising.

---

[4] Although it is possible for users to create an account using a pseudonym, the high frequency with which these information control mechanisms are used (in approximately 50% of contribution instances) indicates that a majority of users reveal their true identity in their user profile.

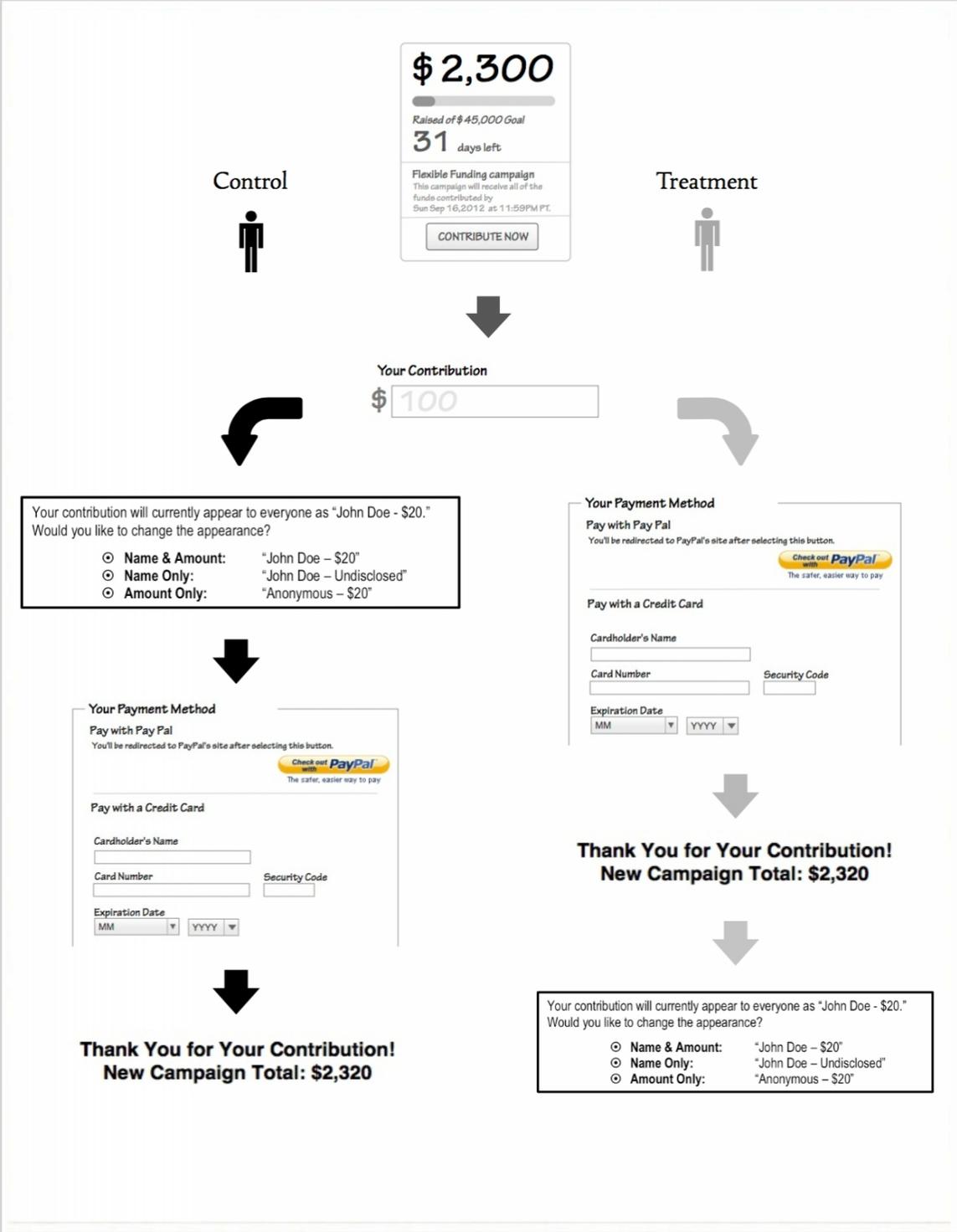

**Figure 2. Contribution Flow & Information Control Prompt Positioning**

This treatment allows us to gain insights into the economic impacts of providing information control mechanisms. Because we only delay the presentation of the mechanism and do not remove it entirely, any identified effects are presumed to be conservative estimates of how provision impacts behavior. Moreover, because we cannot ensure that every campaign visit is associated with a first-time contributor[5], some subjects in our treatment condition may anticipate the eventual provision of information hiding mechanisms. Such anticipatory behavior can only mute the effects of our treatment, again resulting in conservative estimates.

## 2.3 Econometric Specification

Our estimation approach relies primarily on ordinary least squares with campaign fixed effects. All of our estimations additionally incorporate time fixed effects (both in terms of the absolute day on which the observation took place, as well as the day of week), and a variety of other control variables[6] pertaining to both the contributor and the campaign.

We estimate our models in a stepwise fashion, beginning with a simple model that includes only our treatment indicator, *Treatment (T)*, as well as campaign and time fixed effects. We then incrementally incorporate the other controls, namely the funds raised by the campaign to date, *Campaign Balance*, the number of days of elapsed fundraising, *Campaign Days*, a binary indicator of whether the visitor arrived on a mobile device, *User Mobile*, as well as indicators for his or her browser type, *User Browser*, language, *User Language*, and country (based on IP-address), *User Country*. Equation 1 captures our econometric specification.

$$Conversion_{ijt} = \beta * T_{ijt} + \gamma * X_{jt} + \lambda * Z_{it} + \varphi_j + \omega_t + \varepsilon_{ijt} \quad (1)$$

We index users with *i*, campaigns with *j*, and time, in days, with *t*. The coefficient of interest is *β*, capturing the effect of our treatment on conversion rates. *X* is a vector of dynamic campaign controls for fundraising and duration, *Z* is a vector of user / visit controls, including browser, language, country and device, *φ* is a vector of campaign fixed effects, *ω* is a vector of day and day of week fixed effects and, lastly, *ε* is our error term.

---

[5] We offer one robustness check in which we examine our treatment's effect on conditional contribution amongst only new users (i.e., those registering in the 24 hours prior). We can be reasonably sure that recent joiners are first time contributors, and thus are unlikely to hold any prior expectations about the availability of information control features. Our results remain consistent in this estimation.

[6] We do not incorporate user fixed effects because we are unable to identify users who do not contribute any funds. This is not a major concern, however, because our treatment is randomized, and thus extremely unlikely to be correlated with omitted variables. Our estimations also demonstrate that incorporating the various controls at our disposal does little to influence the magnitude or significance of our treatment effect estimates.

We employ a similar specification to estimate our treatment's effect on conditional and unconditional contribution. A notable difference in our conditional contribution estimations, however, is that we are able to identify all subjects in the sample. Accordingly, we can incorporate additional contributor controls associated with the user account, such as his or her tenure on the platform, *User Tenure*, and an indicator of whether he or she has an explicit organizer relationship with the campaign, *Organizer*. Equation 2 captures our specification for the conditional conversion model. Our estimations considering contribution per visitor (unconditional contribution) are identical, except that they exclude the account-based contributor controls.

$$Contribution_{ijt} = \beta * T_{ijt} + \gamma * X_{jt} + \lambda * Z_{it} + \varphi_j + \omega_t + \varepsilon_{ijt} \quad (2)$$

In addition to providing our main regression results, we offer a set of ancillary analyses intended to explore and validate the mechanism underlying our treatment effect. Further, we provide a series of robustness checks – e.g., alternative estimators, sample splits, and manipulation checks.

## 2.4 Data & Descriptive Statistics

Our experiment was conducted over a 14-day period. We observed 128,701 visitors that entered the campaign contribution flow, and thus joined our subject pool. Of these, 62,332 were assigned to the treatment condition (48.4%) and 37,328 chose to contribute funds (29%). The distribution of subjects entering each condition, over the course of our experiment, is presented in Figure 3. Table 1 provides a breakdown of notable descriptive statistics across each stage of the contribution flow, across conditions. Table 2 provides sample-wide descriptive statistics for all of our variables.

-- **INSERT TABLE 1 HERE** --

-- **INSERT TABLE 2 HERE** --

Figures 4, 5 and 6 respectively depict differences in the probability of conversion, expected conditional contribution and expected unconditional contribution between our control and treatment

groups. In each case, we see rather stark shifts in user behavior, with conversion rates increasing and conditional average contributions decreasing.

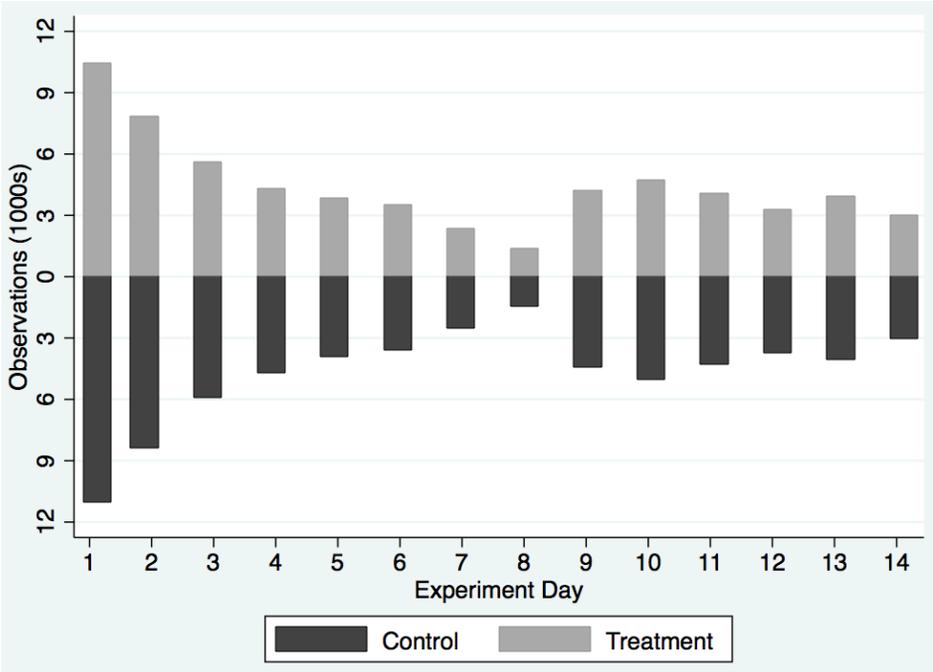

**Figure 3. Treatment and Control Proportions Over Time**

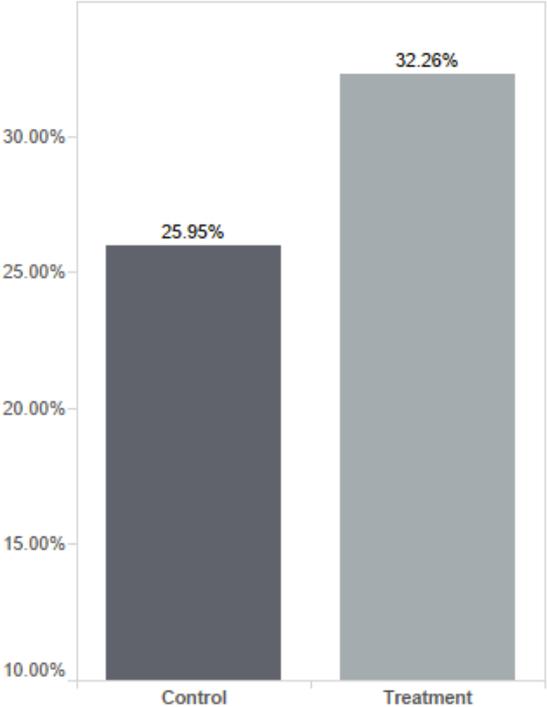

**Figure 4. Conversion Rate between Control and Treatment Groups**

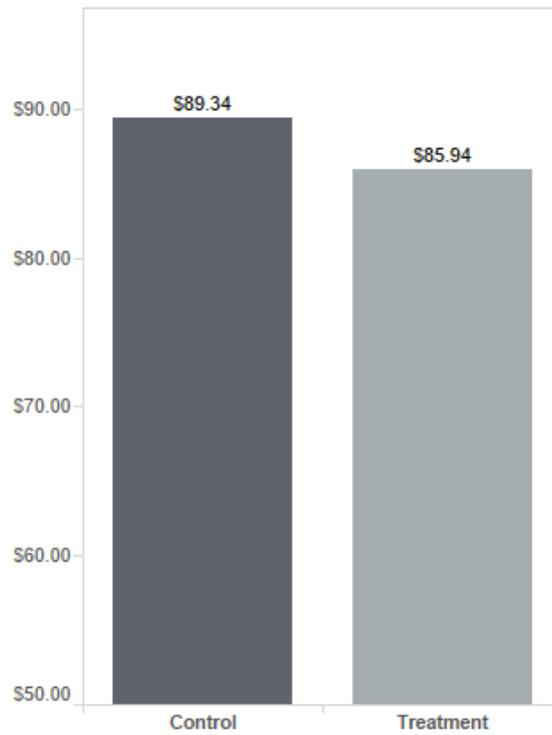

**Figure 5. Average Conditional Contribution between Control and Treatment Groups**

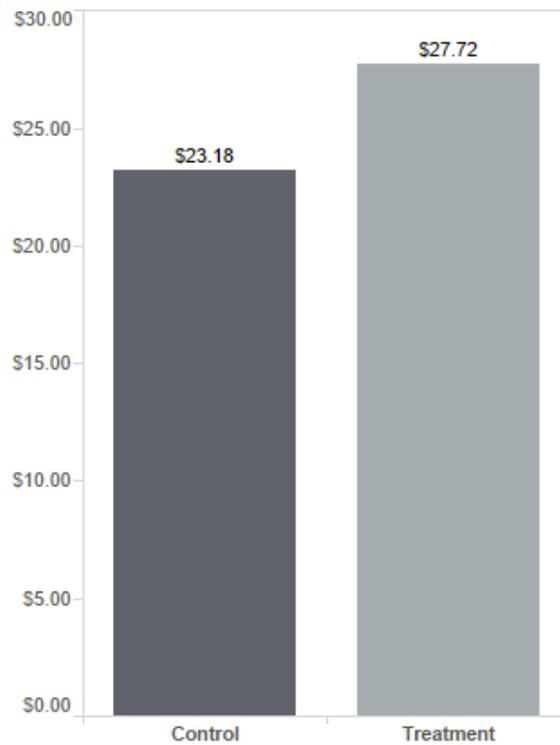

**Figure 6. Average Unconditional Contribution between Control and Treatment Groups**

We also collected additional data about the prevalence with which campaign visitors view prior others' contributions. We obtained data from the platform operator about user navigation patterns. Specifically, we obtained data for roughly 145,000 campaign visitors about the last campaign tab they viewed before navigating elsewhere. We observed that nearly 30% of visitors examined the list of past contributions immediately before navigating elsewhere (either to contribute or exit). Considering that we only observe the last tab viewed, the proportion of visitors navigating to the funders tab is in fact likely to be much higher than this. This provides clear empirical evidence of the potential role of scrutiny and publicity.

## 3. Results

We began by studying the treatment effect on the probability of visitor contribution. As noted previously, the perception of control over one's information, and thus one's privacy, can have multiple countervailing effects. On the one hand, users' perception of control can result in increased rates of participation, if privacy sensitive users are made more comfortable (Hui et al. 2007; Tsai et al. 2011). On the other hand, prompting users with privacy- or scrutiny-related questions can prime users with privacy concerns, thereby reducing participation (John et al. 2011; Tucker 2014). This latter notion is also supported by recent work that has found that individuals actually place less emphasis on privacy when they are not initially endowed with it (Acquisti et al. 2009).

We also assessed the treatment's impact on users' dollar contributions. A number of studies note that individuals go to great lengths to conceal information when they are concerned about how others will view it (Ariely and Levav 2000; Huberman et al. 2005). Here, individuals may prefer to conceal their contributions if they may be viewed as 'cheap'. Alternatively, large donors might fear drawing attention or unwanted solicitations for future donations from other campaigns. For these reasons, we anticipated that the prominence of the information control prompt would be positively associated with extreme contributions (very small or very large), as cognizance of the option to conceal information was expected to make users more willing to engage in such activity.

We first report results for the impact of our treatment on the probability of campaign contribution (Table 3). We saw that the treatment reduces privacy sensitivity, resulting in an approximate 4.9% increase in the probability of conversion (column 4). Examining the change in dollar contributions, conditional on conversion (Table 4), we found that the average contribution declined by approximately $5.81 (column 5). This result reinforces our earlier observation that offering information controls may have a somewhat complex effect, in that it can have a variety of countervailing impacts. Taken together, the above two results indicate that the provision of information hiding mechanisms, and perhaps privacy controls in general, can have counterintuitive, detrimental impacts on user behavior from the purveyor's standpoint, raising users' concerns, lowering their willingness to transact on the platform[7].

-- INSERT TABLE 3 HERE --

-- INSERT TABLE 4 HERE --

When we consider the above effects in tandem (i.e., the combination of increased participation and reduced contribution), we find that the increase in conversion rates dominated. These results are reported in Table 5. Thus, our treatment ultimately resulted in a net benefit for the platform purveyor in terms of overall fundraising outcomes. We saw an estimated increase of roughly $3.55 in the average contribution per visitor, following treatment.

-- INSERT TABLE 5 HERE --

## 4. Supporting Analyses

We next conducted a set of secondary analyses to understand the underlying mechanisms of the observed effects. We examined whether average contributions were indeed falling because of a decline in the variance of contributions (i.e., fewer extreme contributions), as suspected. Further, we looked for heterogeneity in the treatment effect around sensitive campaign topics. We undertook four additional analyses in this regard.

First, we sought to quantify any shifts in the deviation of contributions relative to the overall campaign average. This reference point is appropriate because the definition of an *extreme* contribution should depend on the characteristics of the campaign being supported, and the social norms surrounding it. We determined the absolute deviation from the average for each contribution record. We then regressed that absolute deviation[8] on our binary indicator of treatment. The results are presented in Table 6. Taking the exponential of our coefficient estimate, we found that the treatment produced an approximate 21% decrease in deviations from the campaign average.

---

[7] At the same time, it should also be noted that the provision of these features could reduce consumer surplus. If opting out of a particular transaction is actually an optimal choice, removing privacy controls and thus privacy priming from the contribution process may actually drive the crowdfunder toward sub-optimal behavior. As such, the treatment may impose some unobserved costs upon crowdfunders.

-- INSERT TABLE 6 HERE --

Second, we examined the total variance in contribution amounts between our treatment and control conditions, identifying a statistically significant decrease (F = 1.059, $p < 0.001$). Additional tests based on Levene's robust test statistic, as well as that proposed by Brown and Fortsythe were similarly significant ($p < 0.001$). This result provides further support for our interpretation of the treatment effect on contribution amounts as deriving largely from subjects' increased perception of publicity.

Third, we examined the degree to which information hiding was associated with larger or smaller contributions (the tails of the distribution), and whether the association was balanced between the two. We constructed two binary indicators of contribution size (small or large), based on whether the contribution amount fell into the bottom or top 1%, 5% or 10% of the overall distribution, respectively. We then ran three regressions, modeling a binary indicator of information hiding as a function of each pair of indicators, in addition to our various controls. We obtained the results reported in Table 7.

We saw that contributions at either tail are significantly more likely to be associated with information hiding and we saw an asymmetric effect; larger contributions were almost twice as likely to be associated with information hiding. Moreover, the difference between the two coefficients was statistically significant ($F(1, 3581) = 6.92$, $p < 0.01$).

-- INSERT TABLE 7 HERE --

Fourth, and last, to explore whether our treatment effect varied with the sensitivity of the campaign topic, we constructed an indicator of topic sensitivity and interacted it with our treatment indicator. We first examined the list of campaign categories, of which there were 24. Amongst these, we identified 4 categories that were potentially quite sensitive, where individuals feelings and opinions are somewhat ideological in nature: politics, religion, education and the environment[9]. Based on this, our new indicator variable reflected whether a campaign fell into one of these four categories. The results of this estimation are reported in Table 8 (note: the main effect of campaign type is not identified in this estimation, because the value is static and thus collinear with the fixed effects). We observed that, as anticipated, our treatment effect was much stronger for sensitive campaign topics. Taken together, these results collectively provide support for the notion that publicity plays a central role in our treatment effect.

---

[8] We employ the log of absolute deviation in order to obtain percentage effects. We also include outlier contributions in this estimation, given that such observations contribute in large part to extreme donations in our sample.

-- **INSERT TABLE 8 HERE** --

## 5. Additional Analyses and Alternative Explanations

We also considered alternative explanations for our results. These analyses, as well as the robustness checks that follow, are provided in a supplemental appendix. The most concerning confound for our intervention would be any decline in the complexity of the user interface (UI) resulting from our intervention, and the associated decline in the effort required on the part of the user to complete payment. That is, we might be concerned that the increase in conversion rates was actually due to removal of a radio button from the pre-payment contribution process, which could have simply streamlined the UI. However, this is unlikely to explain the observed effects for a number of reasons.

First, we explored the duration of time it took contributors to complete payment between our treatment and control groups. We found no evidence that the treatment group completed their payments more quickly ($t = -1.26$, $p = 0.21$)[10]. This is important, because we would expect to see significantly shorter visit durations in our treatment group if reduced complexity and effort were to explain our results.

Second, we examined moderating effects associated with visitors' mobile device usage. The UI complexity explanation would suggest that our treatment effect should be amplified for mobile users, who should be more sensitive to UI changes, because of the limitations of smart phone screen size, etc.. However, we find no evidence of a positive moderating effect. This result, reported in Table S1, runs

---

[9] A complete list of campaign categories is provided in the supplementary appendix, in Table S7. Examples of less sensitive topics include Video / Web, Games and Food.

[10] This t-test was performed on logged visit duration in order to meet the assumptions of normality. This analysis also excluded outlier observations in terms of visit durations, namely visits in excess of 1,500 seconds or 25 minutes. We exclude these observations because they likely represent visits where a browser window was left open and inactive.

directly counter to a UI complexity explanation[11].

-- INSERT TABLE S1 HERE --

Third, and last, it is important to keep in mind that UI complexity is completely incapable of explaining the significant decline we observe in average contribution amounts with treatment. Taken in tandem, the above analyses and this last notable fact make it unlikely that UI complexity can explain our findings.

We next sought to delve deeper into the publicity effect. We considered that campaign organizers might contribute to their own campaigns, which we refer to as self-contribution. Noting this, it is conceivable that the contribution effect we observed was largely attributable to campaign organizers ceasing self-contribution in the face of publicity. To assess this, we constructed a binary indicator of self-contribution, and regressed it on our treatment indicator, as well as our set of control variables. If our results were driven by campaign organizers' ceasing self-contribution, then we would expect our treatment indicator to have a significant, negative effect on the probability of any contribution being made by a campaign organizer. As can be seen in Table S2, we observed no evidence of this. It therefore seems unlikely that our results are due to a decline in the rate of organizers supporting their own campaigns.

-- INSERT TABLE S2 HERE --

## 6. Robustness Checks

We explored the robustness of our results in a number of ways. First, we considered the impact of outlier observations. We repeated our primary estimations excluding observations that fell within the top 5% of the distribution in terms of contribution amounts. We also repeated our estimations excluding observations associated with campaigns in the top 5% of the distribution of funding targets. Our results remained generally unchanged in both cases.

Next, we considered the use of alternative estimators. We explored both the conditional Logit and Probit estimators for our conversion model, and we considered fixed effects Poisson and negative binomial estimators for our contribution models. The results of the additional estimations for the treatment's effect on conversion are provided in Table S3. Similarly, the results we obtained using Poisson and Negative Binomial estimators, for our conditional and unconditional contribution models are reported in Tables S4 and S5, respectively. In each case we report marginal effects. In all three cases, we see results that are consistent with those reported in our primary results.

---

[11] Note: we do observe a decline in visit durations but they are not severe enough to produce statistical significance. We provide clustered histograms of logged visit durations comparing the treatment and control groups, as well as comparing mobile and desktop users, in the supplementary appendix, in Figures S1 and S2.

-- INSERT TABLE S3 HERE --

-- INSERT TABLE S4 HERE --

-- INSERT TABLE S5 HERE --

We then re-ran our estimation using a subsample of our data, focusing only on converted visits, amongst users who registered on the platform within the prior 24 hours[12]. The logic here was that new users should be unlikely to hold any expectations about the availability of information controls on the platform, and they should therefore be less likely to notice any changes in the website design. Repeating our conditional contribution estimation on this subsample of observations, we obtained the results reported in Table S6, which exhibit a roughly equivalent treatment effect. We can therefore be confident that our results are not driven by subjects' awareness of alternative conditions.

-- INSERT TABLE S6 HERE --

As a final validation of our results, we considered possible sources of heterogeneity in the treatment effect on conversion. First, we examined possible differences across campaign types that draw different average contribution amounts. We began by calculating average contribution amounts for each campaign type. We then constructed an indicator variable capturing whether a campaign was a "high-spend" category or not, based on whether the campaign was in the top half of this list. We then re-estimated our LPM, incorporating an interaction between the high-spend indicator and our treatment indicator. Doing so, we found no significant effects. We then repeated this process based on median campaign contribution size, and again observed no significant differences[13].

---

[12] Because we can identify everyone, we are able to comprehensively determine the date on which they joined the platform.

## 7. Manipulation Checks

Following the above, we undertook a manipulation check for our intervention, assessing shifts in the pattern of information hiding mechanism usage between the pre- and post-payment conditions. Logically, delaying access to the information hiding mechanism should drive a reduction in its use if our intervention is having the anticipated effect. As such, we looked for a general downward shift in the mechanism's usage in our treatment condition. As anticipated, the rate of information hiding was found to be much lower, indicating that our treatment did indeed have the desired effect. In particular, in the control condition approximately 47% of contributions involved information hiding, as compared to the treatment condition, where approximately 21% of contributions involved information hiding. These results are depicted graphically in Figure 7.

We next examined whether information hiding (and our treatment's effect on information hiding) depended on campaign characteristics. In order to examine this, we constructed campaign category dummies and interacted them with our treatment indicator. We then regressed a binary measure of information hiding on these various dummy interactions (note: we employed fixed effect estimators, thus the main effects of campaign type were not identified in this estimation).

We found that our treatment had a large, highly negative effect on hiding behavior, as we would expect from the model-free results above ($\beta = -0.279$, $p < 0.001$). However, we found no evidence that the effect was moderated by campaign category, with one exception: the Video and Web category, where the treatment effect was significantly attenuated ($\beta = 0.106$, $p < 0.01$). Our suspicion is that this is because the baseline level of information hiding is already quite low for contributions toward projects in this category, thus the potential impact of the treatment is much lower to begin with. In particular, the rate of information hiding in the Video and Web category in the control condition is 0.33, yet the rate is 0.48 amongst all other categories. In fact, the next lowest rate is 0.41, in the Theatre category.

---

[13] We also examined whether the treatment effect was attenuated when subjects arrived following an anonymous contributor (e.g., if such subjects anticipated eventual access to information controls, even when that access was delayed). However, we found no evidence of this.

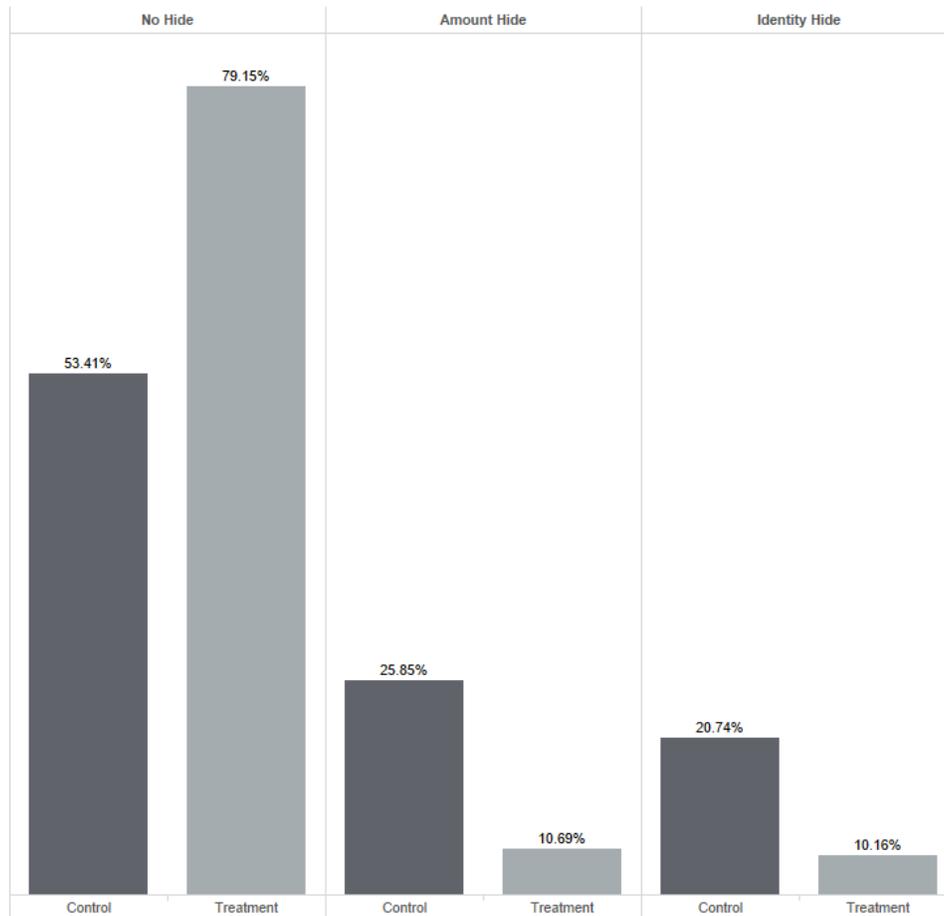

**Figure 7. Probability of Information Hiding by Type (N = 37,328)**

Next, we considered potential interactions between our treatment and the size of the project target. However, we again came to a similar conclusion; the main effect of treatment was comparable to that reported in our category type analysis ($\beta = -0.266$, $p < 0.001$) and the interaction effect, while statistically significant, was extremely small (B = 6.46e-10, p < 0.01). Moreover, when we re-estimated this model replacing project goal with its log, the interaction was completely insignificant. Given these results, it appears that our treatment effect is quite generalizable, and does not depend heavily on the type of campaign being supported.

## 8. Managerial Implications

Our findings indicate that the results of past field work might not tell the entire story when it comes to the impacts of privacy assurances and information controls on consumer behavior. Although numerous studies in the literature have employed laboratory and field experiments to evaluate these issues, generally reporting that these mechanisms increase customer information sharing and transaction likelihood, it is possible (even likely) that past results cannot account for changes in the volume or composition of the converted population that are likely to arise following modifications to a website interface.

Our results can inform crowdfunding stakeholders in a number of ways. First, the provision of information controls should be considered with care. While it is likely that our results would generalize to other reward- and donation-based crowdfunding platforms, or perhaps even equity-based crowdfunding, this will depend heavily on a number of factors. The degree to which the platform enables transparency, reputation and recognition is likely to be important, for example. Therefore, the design of the platform in this regard should be context-dependent. One key factor to consider is the nature of the campaigns typically funded on a given platform. Potentially controversial campaigns are likely to induce greater cognizance and use of information control features. Firms operating platforms with sensitive content will therefore need to take greater care in the design and implementation of information controls.

Additionally, given that there is an inherent tension between enabling recognition for contributors and avoiding issues of privacy and publicity, platform operators and campaign organizers should consider supplemental approaches to mitigating privacy priming in the presence of information controls. For example, organizers might present privacy seals and other forms of reassurance alongside information control prompts. Campaign organizers might also offer recognition to contributors for large contributions by providing participatory rewards and recognizing contributors for their effort – e.g., awarding a large contributor naming rights to a product or thanking them for their participation on the company website – rather than tying recognition to the transaction itself. Contributors could then maintain obscurity by concealing contribution activity, while still benefitting from recognition.

Campaign organizers might also offer the crowd an opportunity to participate and contribute via other, effort-based avenues. Although some contributors might shy away from public monetary contributions, they might be willing to publicly partake in the campaign on an effort basis instead, by volunteering expertise or ideas. Notably, some platforms provide these options (e.g., Spot.us provides an option to 'Donate Talent' to a campaign).

With regard to crowdfunding contributors, our work reinforces the prior finding in other contexts that individuals are often uncertain of privacy risks, and that these perceptions are largely driven by available cues (John et al. 2011). We have shown that the mere presence of information-related prompts can severely impact conversion rates and platform contributions.

It is also important to discuss potential limitations of our work. A key issue that arises here is around user names and pseudonyms. It could be argued that crowdfunders can simply employ a pseudonym if they are really concerned about being observed. However, empirically, we have seen that more than one third of contributions in our sample involve information hiding. This indicates that many crowdfunders do in fact place value on their user profile.

Moreover, this issue is more complex than it might appear at first glance. If users wish to accrue recognition for their actions, it is in their interest to incorporate aspects of their true identity into their user profile. Even for those users who do not do so, online personas tend to persist across transactions and interactions, and thus can carry their own reputation (Dellarocas 2003). This kind of identity disclosure in online personas has been shown to have significant economic outcomes in electronic markets (Ghose and Ipeirotis 2011; Ghose et al. 2012). Notably, reputation and recognition are both factors that have proven to be quite important in offline venture capital, because high profile and well-regarded investors are better able to drive follow on investment (Hochberg et al. 2007; Sørensen 2007; Sorenson and Stuart 2001). Indeed, recent work in crowdfunding has found that expert contributors play a similarly key role in driving follow-on contribution in some markets (Kim and Viswanathan 2013). Moreover, other recent work has noted the role of campaign organizers' social embeddedness in the crowdfunding, as a driver of fundraiser success (Younkin and Kashkooli 2013), as well as the critical role of indirect reciprocity (Zvilichovsky et al. 2013).

## 9. Conclusion

Online spaces are characterized by increased visibility and traceability, and crowdfunding platforms, in particular, publicly record transactions, which include the identity or dollar amounts of campaign contributions. Financial transactions tend to be sensitive in nature, thus publicity and scrutiny may impede transactions. Bearing in mind these issues of visibility, many crowdfunding platforms offer transaction level information controls, so that contributors can decide what will be made publicly visible about their transactions.

Unfortunately, prompting users with information- and scrutiny-related questions can have detrimental effects. On the one hand, prompts of this sort can prime users with privacy concerns. On the other hand, withholding these features could make privacy conscious users less comfortable. With the above tension in mind, we have examined the effect of transaction-level information controls on the behavior of online crowdfunders. Employing a randomized field experiment, we considered the double-edged sword presented by the provision of these features, during the course of the crowdfunder contribution process. We considered both positive (increased comfort or security) and negative effects (privacy priming). We find that delaying the presentation of these mechanisms increases conversion rates, yet simultaneously lowers average dollar contributions.

Although we provide evidence suggesting that privacy priming and publicity effects drive these outcomes, future work can explore the role of mechanism design, wording and presentation format. It is possible that one or both of these effects would be moderated by specific attributes of the mechanism, such as the wording of the text, the granularity of information hiding options (e.g., providing an additional option of presenting a discretized 'range' of the contribution, such as '$10-$20'), or the positioning of the mechanism in the user interface (Egelman et al. 2009).

It is also important to consider the contextual nature of these results, and the degree to which they would generalize to other, non-crowdfunding contexts. It is possible that our results would not extend to a purchase context, where issues of social capital, reputation, etc. might be less pronounced. Further, in regard to the net positive outcome in contributions that we have observed, although a user is given complete freedom here to specify the size of their contribution, thereby allowing for a shift in the distribution of contributions that can offset the decline in participation we would observe when introducing a privacy control question, in other contexts, engagement or contribution may not be up to the user.

To clarify, if transaction amounts are fixed (e.g., a transaction on Amazon.com that involves a product with a fixed price, or a voting type setup, where voters may issue 1, and only 1 vote), then any decline in participation could not be offset by a parallel increase in contributions amounts. In that scenario, the impact of our intervention on participation and unconditional contribution would be strictly negative, as a matter of course. This point highlights the fact that the impact of privacy control provision on user participation and contribution is contextual in a number of different respects, which need to be evaluated in tandem.

Our work shows the potential of large scale *in vivo* randomized experiments to robustly estimate treatment effects around online user behavior, circumventing numerous threats to validity. The methods themselves are widely applicable to research in online contexts, which has ever-increasing relevance and practicality for numerous fields of study. Indeed, given the plethora of influences and information sources available to users in online settings, the complex, messy nature of these contexts means that endogeneity of effects grows increasingly likely. Randomized experiments thus appear to be the best course of action in achieving causal inference, going forward.

# TABLES

**Table 1.**
**Descriptive Statistics: Control vs. Treatment**

| Variable | Control | | | Treatment | | | Tot. N |
|---|---|---|---|---|---|---|---|
| | Mean | St. Dev. | N | Mean | St. Dev. | N | |
| Conversion | 0.259 | 0.438 | 66,369 | 0.323 | 0.467 | 62,332 | 128,701 |
| Conditional Contribution | 89.337 | 260.948 | 17,222 | 85.939 | 263.440 | 20,106 | 37,328 |
| Organizer | 0.017 | 0.128 | 17,222 | 0.019 | 0.136 | 20,106 | 37,328 |
| Binary Hide | 0.470 | 0.500 | 17,222 | 0.208 | 0.406 | 20,106 | 37,328 |

**Table 2.**
**Descriptive Statistics: Sample-Wide**

| Variable | Mean | St. Dev. | Min | Max | N |
|---|---|---|---|---|---|
| Treatment | 0.48 | 0.50 | 0.00 | 1.00 | 128,701 |
| Conversion | 0.29 | 0.45 | 0.00 | 1.00 | 128,701 |
| Unconditional Contribution | 25.38 | 146.73 | 0.00 | 10,000.00 | 128,701 |
| Campaign Balance | 148,134.60 | 253,854.80 | 0.00 | 1,142,523.00 | 128,701 |
| Campaign Days | 25.77 | 22.74 | 1.00 | 252.00 | 128,701 |
| User Mobile | 0.22 | 0.42 | 0.00 | 1.00 | 128,701 |
| User Organizer$^x$ | 0.02 | 0.13 | 0.00 | 1.00 | 37,328 |
| User Tenure$^x$ | 43.15 | 127.49 | 0.00 | 1,835.00 | 37,328 |
| **Day of Week** | | | | | |
| Monday | 0.12 | 0.32 | 0.00 | 1.00 | 128,701 |
| Tuesday | 0.08 | 0.28 | 0.00 | 1.00 | 128,701 |
| Wednesday | 0.19 | 0.39 | 0.00 | 1.00 | 128,701 |
| Thursday | 0.19 | 0.39 | 0.00 | 1.00 | 128,701 |
| Friday | 0.17 | 0.37 | 0.00 | 1.00 | 128,701 |
| Saturday | 0.14 | 0.34 | 0.00 | 1.00 | 128,701 |

*Notes:* x – sample based only on converted users.

**Table 3.**
**Regression Results: Conversion Rate**
**(Linear Probability Model /w Fixed Effects, DV = Conversion)**

| Explanatory Variable | (1) | (2) | (3) | (4) |
| --- | --- | --- | --- | --- |
| Treatment | 0.057*** (0.007) | 0.057*** (0.007) | 0.055*** (0.007) | 0.049*** (0.007) |
| Campaign Balance | -- | 1.17e-07** (3.77e-08) | 1.25e-07** (3.27e-08) | 2.09e-07*** (1.78e-08) |
| Campaign Days | -- | -0.015*** (0.002) | -0.013*** (0.002) | -0.013*** (0.001) |
| User Mobile | -- | -- | -0.152*** (0.009) | -0.156*** (0.009) |
| User Browser | Not Included | Not Included | Not Included | Included |
| User Language | Not Included | Not Included | Not Included | Included |
| User Country | Not Included | Not Included | Not Included | Included |
| Day of Week Effects | Included | Included | Included | Included |
| Time Effects | Included | Included | Included | Included |
| Campaign Effects | Included | Included | Included | Included |
| Observations | 128,701 | 128,701 | 128,701 | 128,701 |
| F-stat | 27.80 (20, 5077) | 33.33 (22, 5077) | 57.78 (23, 5077) | 1.2e+08 (214, 5077) |
| R-square | 0.14 | 0.14 | 0.15 | 0.18 |

*Notes:*

1. *** $p < 0.001$, ** $p < 0.01$.
2. *Robust standard errors reported in brackets for coefficients, clustered by campaign.*
3. *Sample includes all users who entered the contribution flow.*

**Table 4.**
**Regression Results: Conditional Contribution**
**(Ordinary Least Squares /w Fixed Effects, DV = Contribution)**

| Explanatory Variable | (1) | (2) | (3) | (4) | (5) |
|---|---|---|---|---|---|
| Treatment | **-5.472* (2.727)** | **-5.472* (2.727)** | **-5.472* (2.726)** | **-5.525* (2.720)** | **-5.810* (2.679)** |
| Campaign Balance | -- | 4.22e-06 (0.000) | 4.22e-06 (0.000) | 5.21e-06 (0.000) | 7.79e-07 (0.000) |
| Campaign Days | -- | 0.703 (1.115) | 0.703 (1.108) | 0.617 (1.108) | **-2.954+ (1.781)** |
| User Mobile | -- | -- | -0.064 (4.148) | 0.458 (4.152) | -3.095 (5.772) |
| User Tenure | -- | -- | -- | **-0.024** (0.009)** | **-0.020* (0.008)** |
| User Organizer | -- | -- | -- | **81.792** (25.129)** | **82.868** (25.184)** |
| User Browser | Not Included | Not Included | Not Included | Not Included | Included |
| User Language | Not Included | Not Included | Not Included | Not Included | Included |
| User Country | Not Included | Not Included | Not Included | Not Included | Included |
| Day of Week Effects | Included | Included | Included | Included | Included |
| Time Effects | Included | Included | Included | Included | Included |
| Campaign Effects | Included | Included | Included | Included | Included |
| Observations | 37,328 | 37,328 | 37,328 | 37,328 | 37,328 |
| F-stat | **1.79 (20, 3581)** | **2.20 (21, 3581)** | **2.06 (23, 3581)** | **2.73 (25, 3581)** | **1.2e+09 (216, 3581)** |
| R-square | 0.17 | 0.17 | 0.17 | 0.17 | 0.18 |

*Notes:*

1. ** $p < 0.01$, * $p < 0.05$, + $p < 0.10$.
2. *Robust standard errors reported in brackets for coefficients, clustered by campaign.*
3. *Sample includes only converted users (i.e., those who contributed at least some amount of money).*
4. *Estimation includes additional user-profile specific controls: User Tenure and User Mobile, because all users are identified.*

**Table 5.**
**Regression Results: Unconditional Contribution**
**(Ordinary Least Squares /w Fixed Effects, DV = Contribution)**

| Explanatory Variable | (1) | (2) | (3) | (4) |
|---|---|---|---|---|
| Treatment | **4.375*** (1.015)** | **4.396*** (1.022)** | **4.232*** (1.003)** | **3.552*** (0.953)** |
| Campaign Balance | -- | **3.01e-05*** (7.34e-06)** | **3.10e-05*** (7.10e-06)** | **3.83e-05*** (6.23e-06)** |
| Campaign Days | -- | **-1.498** (0.481)** | **-1.344** (0.467)** | **-1.300** (0.443)** |
| User Mobile | -- | -- | **-15.131*** (2.482)** | **-16.386*** (2.604)** |
| User Browser | Not Included | Not Included | Not Included | Included |
| User Language | Not Included | Not Included | Not Included | Included |
| User Country | Not Included | Not Included | Not Included | Included |
| Day of Week Effects | Included | Included | Included | Included |
| Time Effects | Included | Included | Included | Included |
| Campaign Effects | Included | Included | Included | Included |
| Observations | 128,701 | 128,701 | 128,701 | 128,701 |
| F-stat | **3.38 (20, 5077)** | **5.41 (22, 5077)** | **6.75 (23, 5077)** | **6.5e+08 (214, 5077)** |
| R-square | 0.06 | 0.06 | 0.06 | 0.06 |

*Notes:*

1. *** $p < 0.001$, ** $p < 0.01$.
2. *Robust standard errors reported in brackets for coefficients, clustered by campaign.*
3. *Sample includes all users who entered the contribution flow.*

**Table 6.**
**Regression Results: Publicity Effect**
**(DV** = Log(Absolute(Deviation))**)**

| Explanatory Variable | OLS-FE |
|---|---|
| Treatment | **-0.192*** (0.044)** |
| Controls[3] | Included |
| Observations | 33,746 |
| F-stat | **7.5e+09 (216, 2517)** |
| $R^2$ | 0.05 |

*Notes:*

1. *** $p < 0.001$.
2. *Robust standard errors reported in brackets for coefficients, clustered by campaign.*
3. *Estimation incorporates the same set of controls used in Table 4, column 5.*
4. *Sample includes all observations that resulted in contribution, with the exception of those that arrived to a campaign first (i.e., first contribution in the sequence). Accordingly, sample only includes campaigns that received more than one contribution.*

**Table 7.**
**Regression Results: Contribution Size and Info Hiding**
**(Linear Probability Model /w Fixed Effects, DV** = Binary Info Hiding**)**

| Explanatory Variable | (1 - 10%) | (2 - 5%) | (3 - 1%) |
|---|---|---|---|
| Large | **0.070*** (0.010)** | **0.128*** (0.012)** | **0.153*** (0.026)** |
| Small | **0.033*** (0.009)** | **0.043** (0.014)** | **0.075** (0.027)** |
| Controls[3] | Included | Included | Included |
| Observations | 37,328 | 37,328 | 37,328 |
| F-stat | **2,022.83 (216, 3581)** | **3.2e+09 (216, 3581)** | **1.4+e09 (216, 3581)** |
| R-square | 0.21 | 0.21 | 0.21 |

*Notes:*

1. *** $p < 0.001$; ** $p < 0.01$.
2. *Robust standard errors reported in brackets for coefficients, clustered by campaign.*
3. *Estimation incorporates treatment indicator, campaign-level fixed effects, day fixed effects, day of week fixed effects, browser language effects, browser type effects, etc.*
4. *Sample includes all converted visitors (i.e., those who contributed at least some amount of money).*

**Table 8.**
**Regression Results: Topic Sensitivity**
**(Linear Probability Model /w Fixed Effects, DV = Conversion)**

| **Explanatory Variable** | **(1)** |
|---|:---:|
| Treatment | **0.044*** (0.007)** |
| Treatment X Sensitive | **0.090*** (0.033)** |
| Controls[3] | Included |
| Observations | 128,701 |
| F-stat | **1.2e+08 (212, 5077)** |
| R-square | 0.17 |

*Notes:*

1. *** $p < 0.001$.
2. *Robust standard errors reported in brackets for coefficients, clustered by campaign.*
3. *Estimation includes the same controls used in Table 3, column 4.*
4. *Sample includes all visitors who entered the contribution flow.*

# SUPPLEMENTARY APPENDIX

## Table S1.
### Alternative Explanations: Mobile Interaction
### (Linear Probability Model /w Fixed Effects, DV = Conversion)

| Explanatory Variable | (1) | (2) |
| --- | --- | --- |
| Treatment | **0.049*** (0.007)** | **0.054*** (0.007)** |
| Mobile User | **-0.156*** (0.009)** | **-0.145*** (0.011)** |
| Treatment X Mobile User | -- | **-0.023*** (0.006)** |
| Controls[3] | Included | Included |
| Observations | 128,701 | 128,701 |
| F-stat | **3.3+e08 (215, 5077)** | **2.4+e08 (215, 5077)** |
| $R^2$ | 0.18 | 0.18 |

*Notes:*

1. *** $p < 0.001$; ** $p < 0.01$; * $p < 0.05$.
2. *Robust standard errors reported in brackets for coefficients, clustered by campaign.*
3. *Estimation incorporates the same set of controls used in Table 3, column 4.*
4. *Sample includes all users who entered the contribution flow.*

## Table S2.
### Alternative Explanations: Self-Contribution
### (Linear Probability Model /w Fixed Effects, DV = Organizer)

| Explanatory Variable | (1) |
| --- | --- |
| Treatment | 0.001 (0.002) |
| Controls[3] | Included |
| Observations | 37,328 |
| F-stat | **6.3e+06 (214, 3581)** |
| R-square | 0.38 |

*Notes:*

1. *** $p < 0.001$; ** $p < 0.01$; * $p < 0.05$.
2. *Robust standard errors reported in brackets for coefficients, clustered by campaign.*
3. *Estimation incorporates the same set of controls used in Table 4, column 5, with the exception of User Organizer and Tenure.*
4. *Sample includes all converted visitors (i.e., those who chose to give some amount of money).*

**Table S3.**
**Robustness Checks: Conversion Rate**
**(DV = Conversion)**

| Explanatory Variable | PROBIT | LOGIT |
|---|---|---|
| Treatment | **0.075*** (0.016)** | **0.077*** (0.016)** |
| Time Effects | Included | Included |
| Campaign Effects | Included | Included |
| Observations | 3,188 | 3,188 |
| Wald Chi-square | **500.66 (112)** | **417.67 (112)** |
| LL | -1,587.58 | -1,587.08 |

*Notes:*

1. *** $p < 0.001$.
2. *Probabilistic marginal effect estimates reported, with bootstrap standard errors in brackets.*
3. *Sample includes all users that entered the contribution flow for a random draw of 100 campaigns. We employ a random subsample for the sake of computational tractability.*

**Table S4.**
**Robustness Checks: Conditional Contribution**
**(DV = Contribution)**

| Explanatory Variable | POISSON | NBREG |
|---|---|---|
| Treatment | **-0.063*** (0.001)** | **-0.056*** (0.008)** |
| Time Effects | Included | Included |
| Campaign Effects | Included | Included |
| Observations | 36,264 | 36,264 |
| Wald Chi-square | **16,307.56 (14)** | **141.43 (14)** |
| LL | -2,050,819.80 | -170,587.25 |

*Notes:*

1. *** $p < 0.001$.
2. *Marginal effect estimates reported, with bootstrap standard errors in brackets.*
3. *Sample includes all converted visitors (i.e., those that contributed at least some amount of money). Note that these estimators drop observations associated with campaigns where only one contribution took place in our period of observation (i.e., no variance in the DV).*

**Table S5.**
**Robustness Checks: Unconditional Contribution**
**(DV = Contribution)**

| Explanatory Variable | POISSON | NBREG |
|---|---|---|
| Treatment | **0.174*** (0.001)** | **0.263*** (0.011)** |
| Time Effects | Included | Included |
| Campaign Effects | Included | Included |
| Observations | 123,521 | 123,521 |
| Wald Chi-square | **64,880.58 (14)** | **1,158.40 (14)** |
| LL | -6,021,683.00 | -257,373.90 |

*Notes:*

1. *** $p < 0.001$.
2. *Marginal effect estimates reported, with bootstrap standard errors in brackets.*
3. *Sample includes all users who entered the contribution flow, except for those observations associated with campaigns where only one visitor arrived, or where no conversions took place (i.e., no variance in the dependent variable).*

**Table S6.**
**Robustness Checks: Recent Joiners**
**(DV = Contribution)**

| Explanatory Variable | OLS |
|---|:---:|
| Treatment | **-8.037* (3.638)** |
| Campaign Balance | **8.94e-05*** (2.33e-05)** |
| Campaign Days | -0.394 (1.258) |
| User Mobile | -4.155 (6.744) |
| User Organizer | 15.371 (23.549) |
| User Browser | Included |
| User Language | Included |
| User Country | Included |
| Day of Week Effects | Included |
| Time Effects | Included |
| Campaign Effects | Included |
| Observations | 24,605 |
| F-stat | **9.7e+09 (215, 3057)** |
| $R^2$ | 0.20 |

*Notes:*

1. *** $p < 0.001$; * $p < 0.05$.
2. Robust standard errors in brackets.
3. Estimation incorporates campaign-level fixed effects, day fixed effects, day of week fixed effects, browser language effects, browser type effects, whether the user is on a mobile device, etc.
4. Sample includes converted users who joined the platform within the prior 24 hours.

**Table S7.**
**Campaign Categories**

| Category | Count |
|---|---|
| Animals | 157 |
| Art | 151 |
| Comic | 39 |
| Community | 760 |
| Dance | 52 |
| Design | 105 |
| Education | 435 |
| Environment | 84 |
| Fashion | 67 |
| Film | 856 |
| Food | 64 |
| Gaming | 130 |
| Medical | 405 |
| Music | 451 |
| Photography | 68 |
| Politics | 29 |
| Religion | 52 |
| Small Business | 281 |
| Sports | 103 |
| Technology | 226 |
| Theatre | 215 |
| Transmedia | 27 |
| Video / Web | 162 |
| Writing | 159 |

*Figures S1 and S2 provide clustered histograms of logged visit duration for treatment vs. control and mobile vs. desktop, respectively. Most notably, we observe that subjects in the treatment and mobile groups exhibit systematically shorter visit durations, though the difference is not statistically significant.*

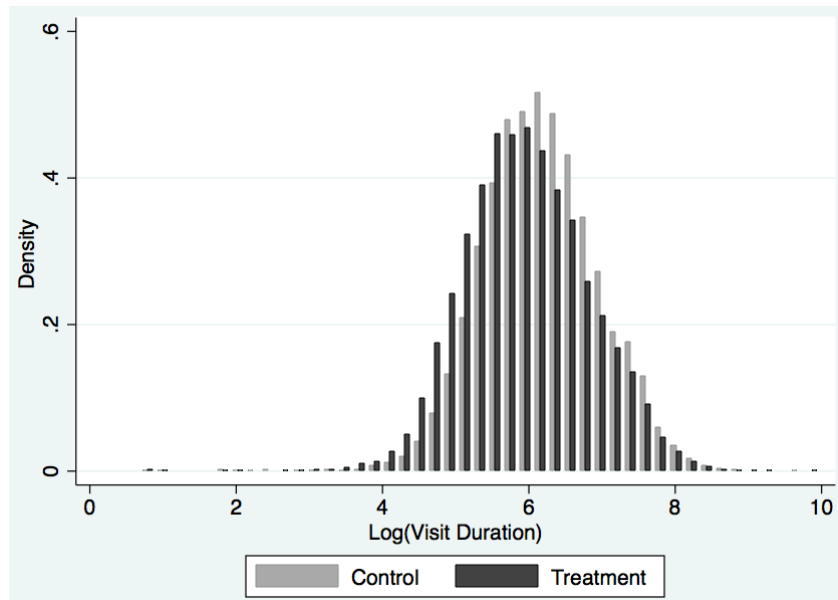

*Figure S1.*
*Histogram of Log(Duration) by Condition (Converted Users)*

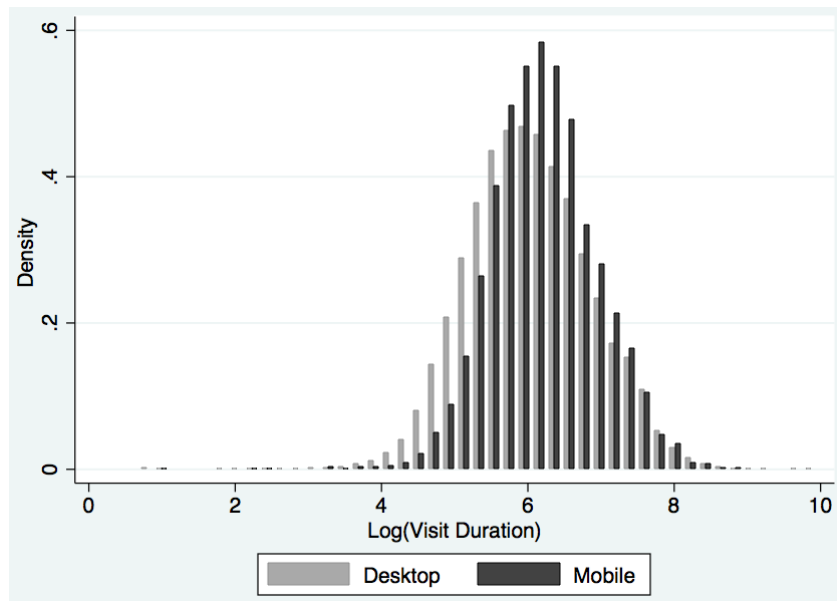

*Figure S2.*
*Histogram of Log(Duration) by Device Type (Converted Users)*